\documentclass[preprint,onecolumn,12pt]{article}
\usepackage{extsizes}
\usepackage[super,sort&compress,comma]{natbib}
\usepackage[version=3]{mhchem}
\usepackage[left=2cm, right=2cm, top=1.785cm, bottom=2.0cm]{geometry}
\usepackage{balance}
\usepackage{times,mathptmx}
\usepackage{sectsty}
\usepackage{graphicx}
\usepackage{lastpage}
\usepackage[format=plain,justification=justified,singlelinecheck=false,font={stretch=1.125,small,sf},labelfont=bf,labelsep=space]{caption}
\usepackage{float}
\usepackage{fancyhdr}
\usepackage{fnpos}
\usepackage[english]{babel}
\addto{\captionsenglish}{%
  
}
\usepackage{array}
\usepackage{droidsans}
\usepackage{charter}
\usepackage[T1]{fontenc}
\usepackage[usenames,dvipsnames]{xcolor}
\usepackage{setspace}
\usepackage[compact]{titlesec}
\usepackage{hyperref}

\usepackage{epstopdf}

\definecolor{cream}{RGB}{222,217,201}

\begin{document}

\title{Synaptic dynamics in complex self-assembled nanoparticle networks}
\author{S. K. Bose,\textit{$^{a}$} S. Shirai,\textit{$^{a}$} J. B. Mallinson,\textit{$^{a}$} and S. A. Brown\textit{$^{a}$}}

\maketitle

\begin{abstract}
We report a detailed study of neuromorphic switching behaviour in inherently complex
percolating networks of self-assembled metal nanoparticles.  We show that
variation of the strength and duration of the electric field applied to
this network of synapse-like atomic switches allows us to control the switching
dynamics. Switching is observed for voltages above a well-defined threshold, with higher voltages leading to increased switching rates. We demonstrate two behavioral archetypes and show how the switching dynamics change as a function of duration and amplitude of the voltage stimulus. We show that the state of each synapse can influence the activity of the other synapses, leading to complex switching dynamics. We further demonstrate the influence of the morphology of the network on the measured device properties, and the constraints imposed by the overall network conductance.  The correlated switching dynamics, device stability over long periods, and the
simplicity of the device fabrication  provide an attractive pathway to
practical implementation of on-chip neuromorphic computing.

\end{abstract}


\renewcommand*\rmdefault{bch}\normalfont\upshape
\rmfamily
\section*{}
\vspace{-1cm}


\footnotetext{\noindent \textit{$^{a}$ The MacDiarmid Institute for Advanced Materials and Nanotechnology, School of Physical and Chemical
Sciences, University of Canterbury, Private Bag 4800, Christchurch 8140, New Zealand. E-mail: simon.brown@canterbury.ac.nz; saurabh.bose@canterbury.ac.nz}
\\ .
\\ This is a post-peer-review, pre-copyedit version of an article published in Faraday Discussions. The final authenticated version is available online at: http://dx.doi.org/10.1039/c8fd00109j}




\section{Introduction}

The quest to replicate the most advanced computer, i.e. the biological brain, has led to immense advancements in software and hardware emulation of the organic neuronal and synaptic structures.\cite{Mead1990a,Merolla2014,Schmidhuber2015}
The central concepts are that nanoscale switching elements, whose
state (resistance) reflects the history of their inputs
(hence the name `memristors')\cite{Strukov2008} can be used to emulate the functions of
biological synapses, and that networks of such elements could
potentially replicate some of the functions of the
brain.\cite{Service2014,Alibart2013,Prezioso2015,Pantazi2016,Stieg2012,Avizienis2012,Serb2016} Recent work has
shown that memristor networks can potentially be used for many
tasks, ranging from simple pattern recognition and associative
learning \cite{Kulkarni2012,Pershin2010} to more complex
applications requiring greater
functionality.\cite{Prezioso2015,Alibart2013,Serb2016,Pantazi2016}

\begin{figure*}[h]
\centering
  \includegraphics[width=12cm]{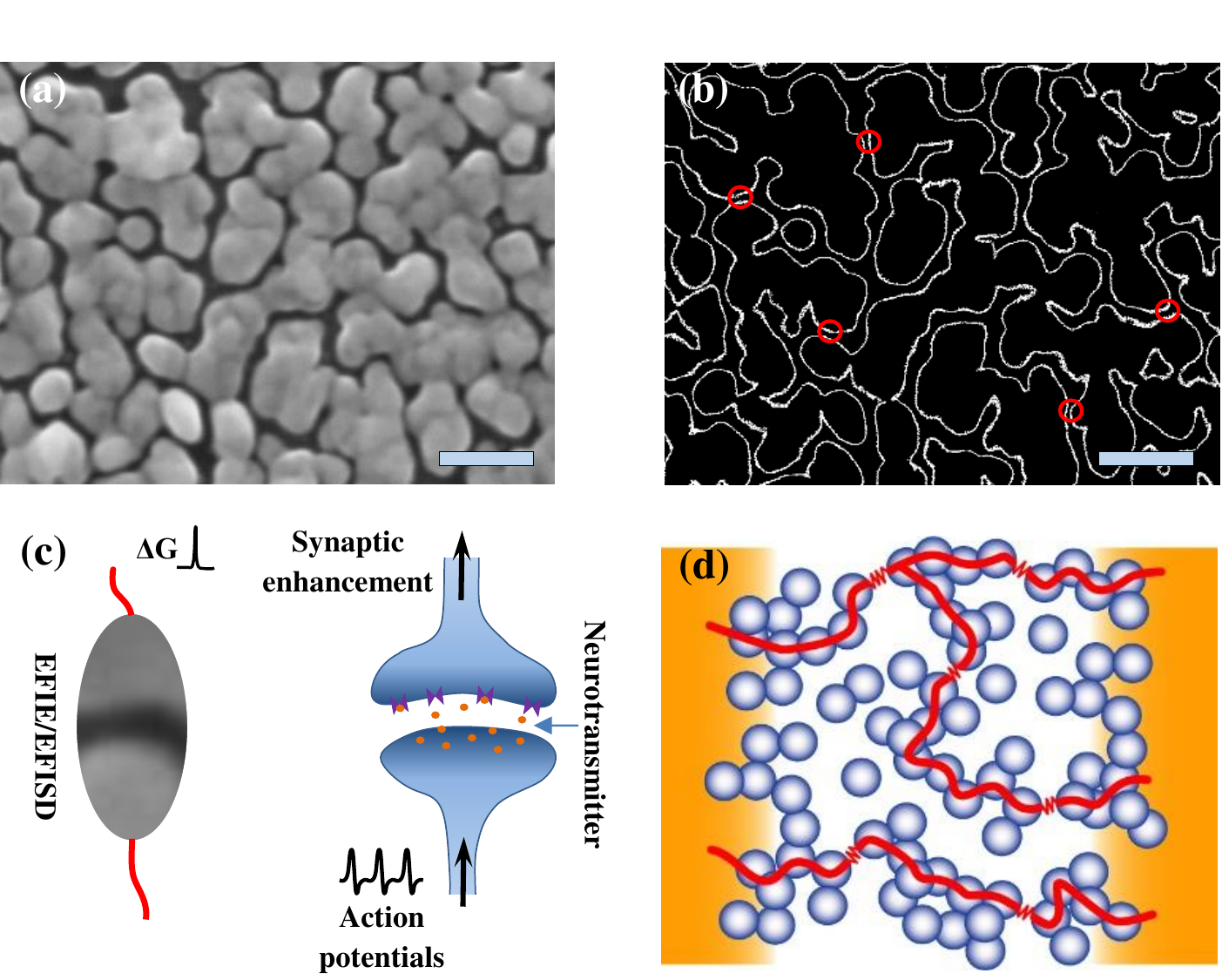}
  \caption{ (a) Scanning electron micrograph
  of a typical part of the nanoparticle network poised near percolation threshold. (b) Corresponding edge-marking
emphasizing the percolating nature of the conduction paths along with the spatially distributed tunnel gaps, which are  potential atomic-wire
formation sites. Some are  highlighted by red circles. The scale bars are
100 nm. (c) On receiving requisite action-potential a biological synapse has a finite probability for enhanced synaptic connection. Similarly voltage-stimulus on individual atomic-switch synapses results in  atomic movements in the tunnel gap via electric-field-induced-evaporation (EFIE)/ electric-field-induced-surface-diffusion (EFISD) processes. There is a finite probability of atomic-wire formation in the tunnel-gap with an associated increase in conductance (measured as $\Delta$G). (d) Such synaptic reconfigurations of atomic-wires results in dynamical activity involving multitudes of interconnected signal pathways. Only a few representative pathways between the electrodes are depicted here.}
  \label{fgr:Fig_Sch2}
\end{figure*}

A wide variety of types of memristive element
\cite{Ohno2011,Kim2012a,Tuma2016} and types of network
\cite{Stieg2012,Avizienis2012,Serb2016} have been proposed, each
of which has strengths and weaknesses. Highly organised arrays of
memristors\cite{Prezioso2015,Wang2017a,Wang2018a} can take advantage of established lithography
techniques, but suffer from the same cost and fabrication
limitations as traditional complementary metal-oxide-semiconductor
(CMOS) devices.\cite{Frank2002,Taur1997} Self-organised
networks\cite{Avizienis2012,Fostner2015} might provide similar
functionality using low-cost fabrication techniques as well as
genuinely neuromorphic \emph{structures} that have the built-in
complexity that is believed to be important for brain-like
functionality.\cite{Chialvo2010,Beggs2003}

Here we focus on networks of metallic nanoparticles which, when poised near the percolation
threshold, exhibit complex switching behaviour due to the formation and annihilation of
atomic scale wires within tunnel gaps in the network\cite{Sattar2013}.  Switching events
that cause increases (decreases) in device conductance
$G_{\uparrow}$($G_{\downarrow}$) are understood\cite{Sattar2013, Minnai2017} to
be due to electric-field-induced formation\cite{Olsen2012}
(electromigration-induced annihilation\cite{Xiang2009}) of
atomic-wires in tunnel gaps within the network. These atomic-switches act as inorganic-synapses and multiple such
switching events (synaptic reconfigurations) in different locations lead to complex switching
dynamics. It has recently been shown that these synaptic networks can be
fabricated such that devices work reproducibly for several
months.\cite{Bose2017}

Since the networks are complex, it is intrinsically difficult to
separate out the switching of the individual elements from the
observed changes of the network as a whole. Here we apply controlled
voltage pulses and characterize the resultant network dynamics. We
show that there is a well-defined threshold voltage ($V_T$) for
the onset of switching, and that the atomic wire formation
processes occur effectively instantaneously on the timescale of
our existing measurement system. Synaptic reconfiguration with voltage stimulus close to $V_T$, where
wire breaking processes occur on longer time scales and individual switching events can be resolved, simplifies the data analysis.
The formation and annihilation of atomic-wires in individual synapses appear to be temporally stochastic, which is believed to
be advantageous for some neuromorphic computation
paradigms\cite{Tuma2016,Querlioz2013}. Each synapse is inter-connected with the complex network and we show here preliminary evidence suggesting heavy tailed distributions of inter-event-intervals (times between switching events), which are characteristic of correlations.\cite{Barabasi1999,Stieg2012} We also show that the
morphology of the network controls its resistance and therefore
shifts the measured conductance away from the quantised
conductance values expected for individual atomic scale
wires.\cite{Agrait2003, Sattar2013} Finally we outline the prospects for further device and applications development.

\section{Experimental}
Our devices are fabricated using simple cluster deposition
techniques that have been described in detail previously
\cite{Schmelzer2002,Bose2017}. Briefly, tin clusters ($\sim$ 8.5
nm) are deposited between Au contacts (100 $\mu$m separation) in a
controlled atmosphere containing a small amount of air ($P_{dep}
\sim 1 \times 10^{-5}$ Torr) with high relative
humidity,\cite{Bose2017} resulting in the morphology shown in
Fig. \ref{fgr:Fig_Sch2} (a). Deposition is terminated when the device
resistance reaches $\sim$2k$\Omega$ ($G \sim 6 G_0$, where $G_0 =
2 e^2/h$ is the quantum of
conduction\cite{Wharam1988,VanWees1988}).

\section{Results I: Network dynamics}
\subsection{Synaptic network of atomic-switches}
\label{network}
Near the percolation threshold, networks of self-assembled nanoparticles contain a multitude of interconnected atomic-switches.\cite{Sattar2013,Bose2017,Fostner2015} Fig. \ref{fgr:Fig_Sch2} (a) shows a typical scanning electron micrograph (SEM) of such a network with percolating pathways. The corresponding edge marking in Fig. \ref{fgr:Fig_Sch2} (b) reveals the presence of gaps in these pathways, some of which are indicated by red-circles. These are potential zones for atomic-wire formation (and annihilation) which can be modelled\cite{Ohno2011} as inorganic counterparts of the organic synapses in the biological brain. The arrival of an appropriate action potential on a biological synapse has a finite probability of enhancing the synaptic connection in Fig. \ref{fgr:Fig_Sch2}(c). In our inorganic-synapse (atomic-switch), the voltage-pulse stimulus can initiate electric-field induced atomic-rearrangements leading to formation of an atomic-wire bridging the gap, and thus enhancing the overall device conductance. Continuous current flow through these wires can reset the synapse via electromigration. The effect of individual synaptic reconfigurations on the measured device conductance i.e. $G_{\uparrow}$($G_{\downarrow}$) is dependent upon the spatial as well as temporal configuration, i.e. the position of the synapse in the network. A very simplistic depiction of such a network is shown in  Fig. \ref{fgr:Fig_Sch2} (d). This complex synaptic network could be imagined as a dynamic reservoir\cite{Mantas2009,Sillin2013,Kulkarni2012} of
interconnected and interdependent synaptic `weights'. Reservoir computing (RC) using such atomic switch networks is believed to be a powerful tool for a range of problems including pattern recognition\cite{Avizienis2012,Mantas2009,Bose2017,Mizrahi2018} and various waveform regression
tasks \cite{Demis2016}.

\begin{figure*}[h]
\centering
  \includegraphics[height=6cm]{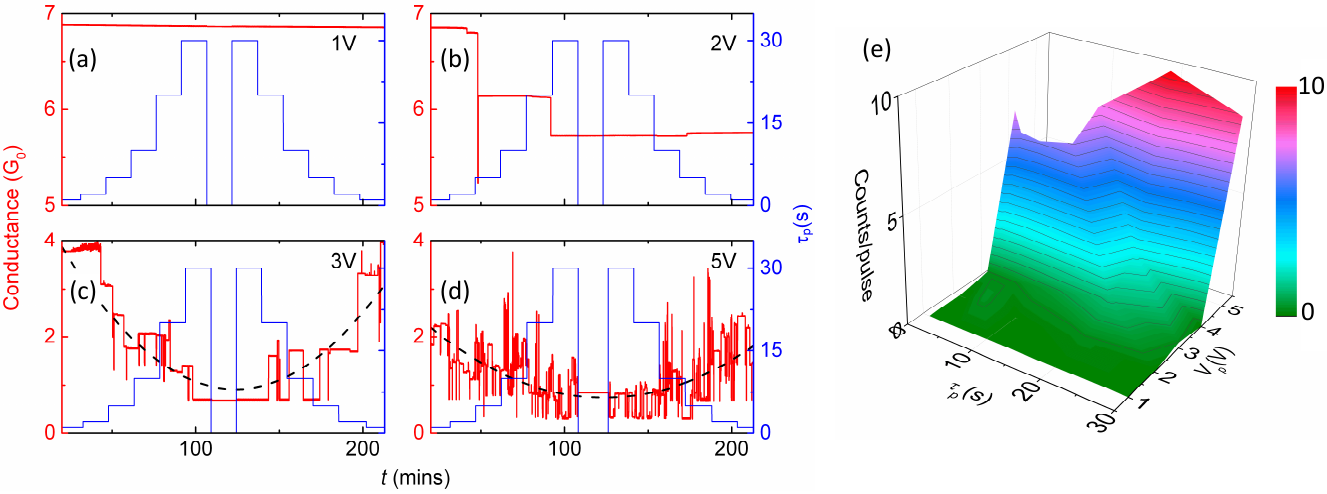}
  \caption{Pulse width $\tau_p$ and voltage $V_p$
dependence of the switching behaviour. (a - d) Standardized pulse
sequence ($\tau_p$ from 2 - 30s) with $V_p$ = 1, 2, 3 and 5V. (e)
Summary plot of the switching rate as a function of both
$\tau_p$ and voltage $V_p$. Note the dramatic increase in switching rate with increasing voltage.}
  \label{fgr:fig_Width1}
\end{figure*}

\subsection{Network activity and event-rates}
\label{Rates}

Excitation of the complex network with voltage pulses initiates synaptic reconfigurations
as shown in Fig. \ref{fgr:fig_Width1}. The variation
of conductance of a typical device in response to our standard
pulse sequence (pulse widths $\tau_p$   from 2s to 30s and back to
2s) and when the pulse amplitude $V_p$ is increased is shown  in
Figs. \ref{fgr:fig_Width1}(a) to (d).  The 1V and 2V pulses result in
few switching events, whereas voltages above $V_T \sim$ 3V
lead to a characteristic dependence on pulse width  (parabolae are guides to the eye)
in which the long pulses result in a lower average $G$. There are absolutely no switching
events for read voltages [$\sim$ 100mV, centre regions in
Fig. \ref{fgr:fig_Width1}(a-d)]: this stability for sub-threshold
voltages ($V_p < V_T$) has been verified for extended periods
(weeks) of time.\cite{Bose2017} Every time a synapse connects/disconnects the electric field and current distribution is rearranged and subsequent synaptic activity is dependent upon the modified network state. As $V_p$ is increased beyond $V_T$ switching becomes more active
and the devices reach a state in which the average conductance is
$\sim 2 G_0$. Active switching around this conductance
value can be observed for many months.\cite{Bose2017}
Fig. \ref{fgr:fig_Width1}(e) summarizes the dependence of the number of
events/pulse  as a function of $\tau_p$ and $V_p$, and emphasises
the dramatic increase in activity of the network for stronger
excitation pulses (higher $V_p$).

\begin{figure}
\centering
  \includegraphics[width=9cm]{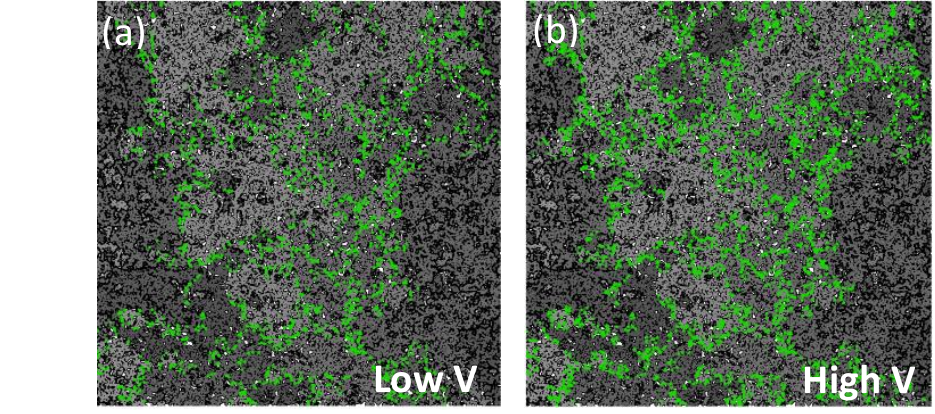}
\caption{Numerical simulation of network activity in a 200 $\times$ 200 network (measured in units of particle size) at low and high voltages. The green markers denote the atomic-switches that reconfigured (active synapses) between the gray shaded ohmically connected groups of nanoparticles. The details about the simulations are discussed elsewhere\cite{Fostner2015}. The increased density as well as spatial distribution of the active synapses at higher voltages (b) indicates activation of previously inactive synapses when low-voltage was applied (a).}
 \label{fgr:fig_sim1}
\end{figure}

This enhanced device activity (higher rate of events) could be due to either increased activity of the \textit{same synapses} or activation of \textit{new previously inactive synapses}. To better understand this, results from numerical simulations\cite{Fostner2015} are shown in Fig. \ref{fgr:fig_sim1}. Similar to experimental measurements, the voltage stimulus is applied on the left side of the network with right side being grounded. Various groups of ohmically connected particles are connected via the active synapses indicated by green markers (i.e. atomic-switches that reconfigured). As can be seen in Fig. \ref{fgr:fig_sim1}(b) the overall density of the active synapses increases at higher voltage which suggests that the experimental observation of increased event-rates at higher voltages can be attributed, at least partially, to the activation of additional synapses. This can be understood intuitively as being due to the increased electric field in tunnel junctions in which the field was previously below a minimum value required to cause atomic motion. In the next section we explore the synaptic dynamics using temperature dependent measurements.

\subsubsection{Temperature dependence of event rates}
\begin{figure}[h]
\centering
  \includegraphics[width=7.5cm]{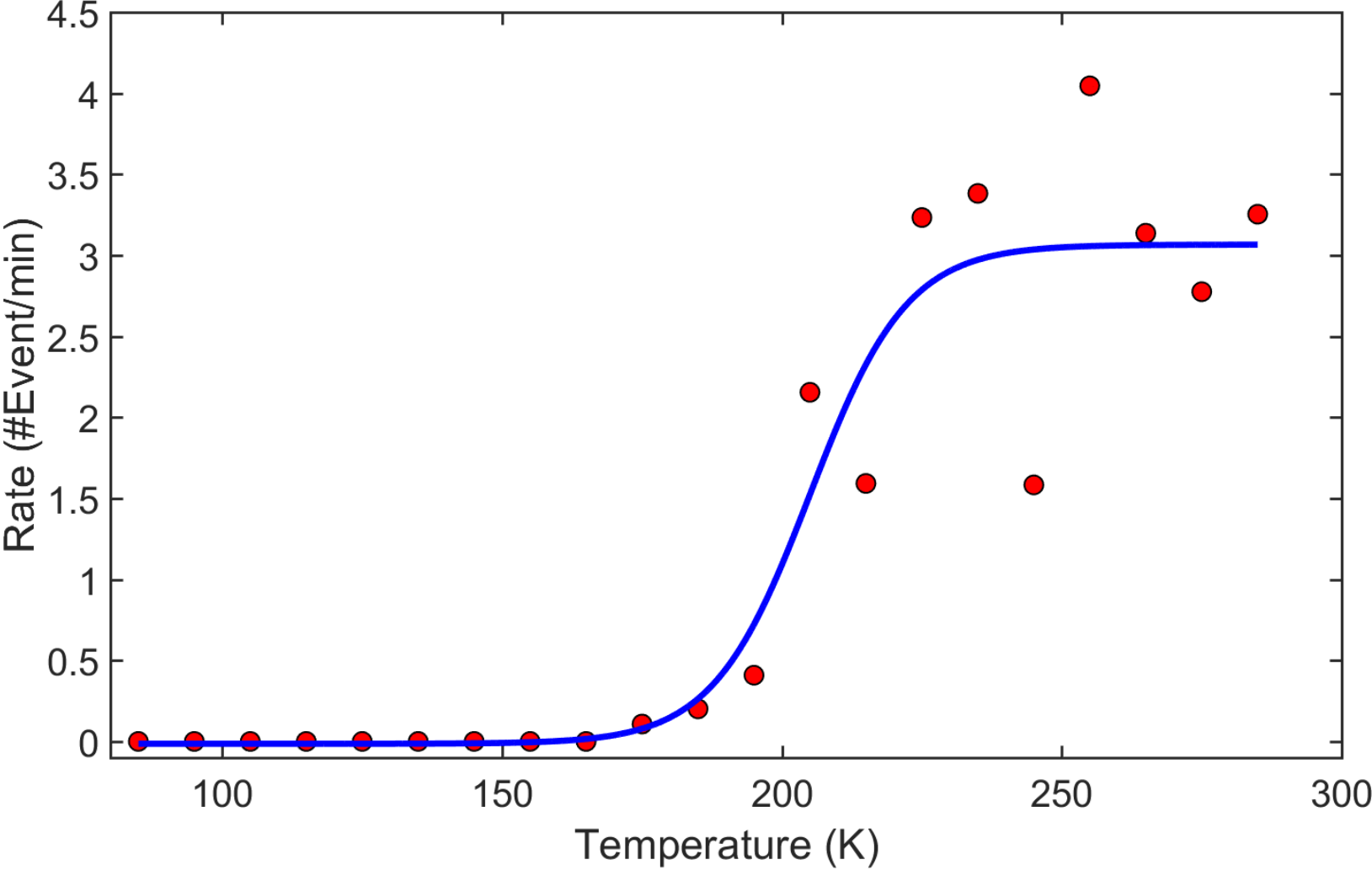}
  \caption{Typical temperature variation of the switching activity, with 8V pulses ($\tau_p = 10s$), shows that there is no switching activity below T$\sim$200K. Switching is activated around 200K and the event rates saturate after 250K. The event rates are averaged every 10K and the solid line is sigmoid-fit.}
  \label{fgr:Temprates}
\end{figure}

The synaptic activity can be arrested by decreasing the device temperature (\textit{T}). As shown in Fig. \ref{fgr:Temprates} lowering the temperature below $T_{th} \sim$ 200K, results in no switching events being observed even when 8V pulses are applied. Application of different voltage pulse amplitudes and pulse-widths (not shown here) leads to similar results. This demonstrates that the typical energy scale necessary for atomic-reconfiguration is $\sim$ 17 meV. This energy scale is very similar to the Debye temperature of Sn\cite{Stewart1983}. Below 200K the atomic motion is `frozen out', effectively preventing  diffusion of adatoms in the tunnel gaps and eliminating the atomic-wire formation processes.

\begin{figure*}[h]
\centering
  \includegraphics[width=11cm]{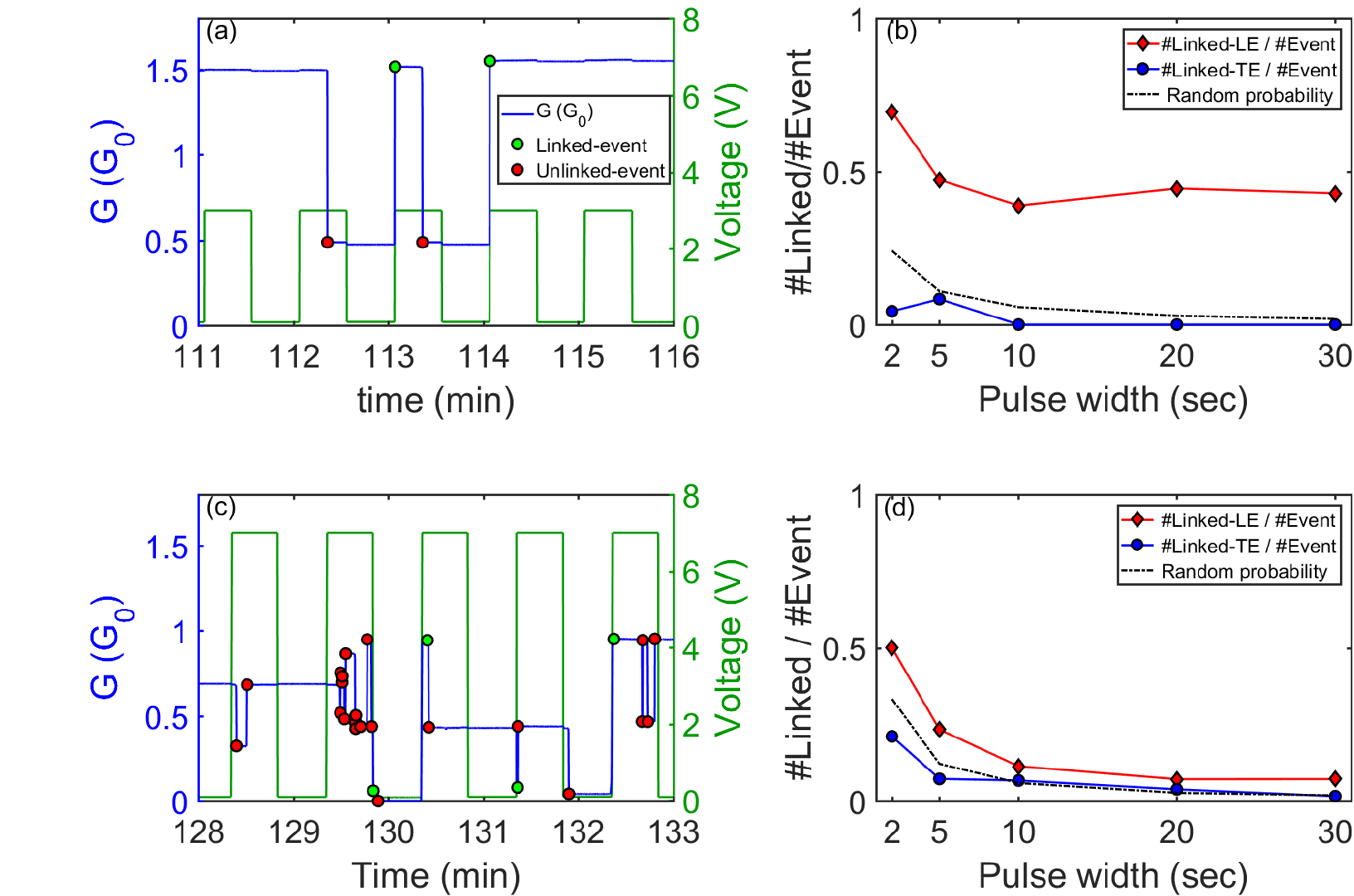}
  \caption{(a, c) Typical snapshot of G(t) traces depicting the events linked and unlinked with the voltage pulse edges for both archetype-A and -B i.e. at low and high-voltages respectively. The green dots mark the linked-events at pulse-edges and red dots mark unlinked-events. (b,d) The pulse-width dependence of the fraction of linked-events that occur at leading (LE) and trailing (TE) pulse edges. Only the LE linked-events are observed to occur more frequently than would be expected from random statistics. The linked-events at the LE are predominantly up-events, e.g. at $\tau_p$ = 30s for archetype A, 92\% linked-events on LE are G$\uparrow$. }
  \label{fgr:Fig_ArchABdep}
\end{figure*}

\subsection{Threshold switching: archetype-A}
\label{archetypeA}
We now explore the switching dynamics in the near threshold regime (V $\sim$ V$_{th}$) where the number of synaptic events is minimized and thus the switching behaviour simplifies. Any voltage pulse stimulus consists of four components i.e. a leading-edge (LE) where voltage increases from off (read voltage 0.1V) to on, trailing-edge (TE) when V decreases from on to off, and on and off regions. Fig. \ref{fgr:Fig_ArchABdep}(a) presents a typical snapshot example of a G(t) trace with $V_p$ close to but above $V_T$. The green markers label events at the pulse-edges as ``linked-events" and the red markers denote events
in the on region as ``unlinked-events". As can be seen in Fig. \ref{fgr:Fig_ArchABdep}(a) for V $\sim$V$_{th}$ the $G_{\uparrow}$ events occur almost exclusively
at the leading edges of the pulses, showing that the electric-field-induced formation of atomic scale wires occurs on a
timescale that is effectively instantaneous\footnote{Ref. \cite{Terabe2005} shows that switching can take place on timescales of nanoseconds.}. In contrast, all $G_{\downarrow}$ events are unlinked, and occur after several seconds of current flow, consistent with the idea that
electromigration breaks the atomic scale wires to reset the synapse. We identify this behavior as archetype-A.

\begin{figure}[h]
\centering
  \includegraphics[height=9cm]{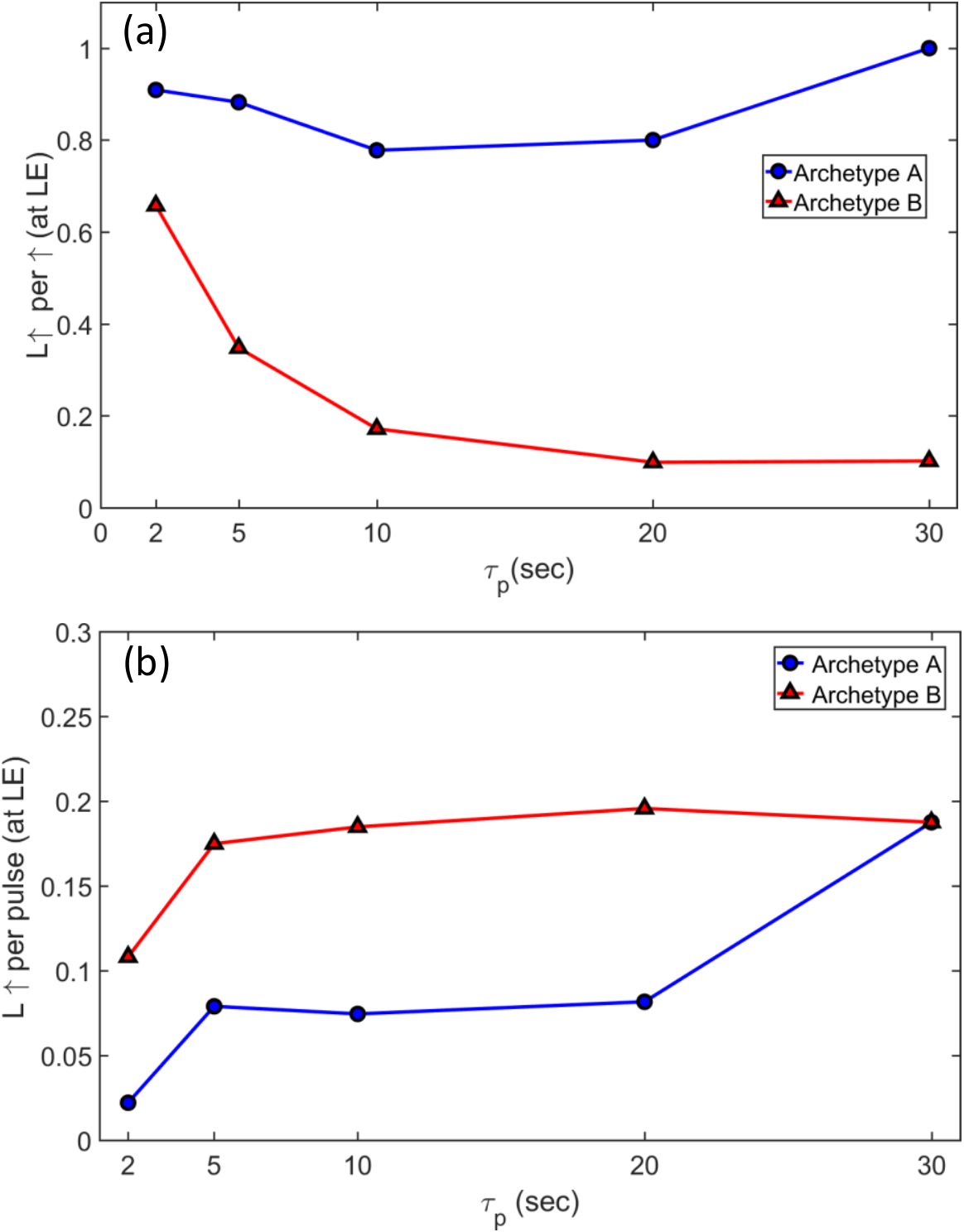}
  \caption{(a) Variation of number of linked-up (L$\uparrow$) events at LE as a fraction of total $\uparrow$ events ($\uparrow = L\uparrow + U\uparrow$) for both archetypes. In archetype-A, L$\uparrow /  \uparrow$ is nearly 1 indicating almost all $\uparrow$ events occur as L$\uparrow$. In archetype-B the L$\uparrow /  \uparrow$ is nearly 0.7 at $\tau_p =$ 2s and decreases as a function of $\tau_p$. This decrease in L$\uparrow /  \uparrow$ for archetype-B is due to additional U$\uparrow$ events (in pairs of U$\uparrow$-U$\downarrow$) events at higher voltages and longer $\tau_p$. (b) The fraction of LE induced L$\uparrow$ events (L$\uparrow /  $ pulse) is in the range around 10-20\% for both archetypes at all $\tau$$_p$ (except low 2s).}
  \label{fgr:Fig_counts}
\end{figure}

Analysis of the switching events over many measurement cycles shows the $G_{\uparrow}$ events are predominantly observed at the leading-edge (LE) of the pulse [Fig. \ref{fgr:Fig_ArchABdep}(b)]. The fraction of events that are linked at the LE (red-diamonds), as a function of the pulse width $\tau$$_p$, is much higher than possible from a random distribution of events (dotted-line). This indicates that a certain proportion of the tunnel gaps in the network respond instantaneously to the applied electric field. In contrast the number of linked-events at the TE is consistent with random occurrence, i.e. these events occur at the TE simply by chance, they are not driven by the changing electric field.

\subsection{V > V$_{th}$: archetype-B}
\label{archetypeB}
At higher voltages i.e. V $> V_{th}$ the archetype-A behavior evolves into more complex archetype-B behavior. Fig. \ref{fgr:Fig_ArchABdep}(c) depicts a typical G(t) snapshot and shows that, in addition to the linked-up events at LE, additional unlinked events (U$\downarrow$ and U$\uparrow$) occur. This behaviour is consistent for all $\tau_p$ and across devices. Similar to archetype-A, the linked-events occur at the LE, rather than the TE.

Fig. \ref{fgr:Fig_counts}(a) shows the variation of the number of linked-up (L$\uparrow$) events at the LE as a fraction of the total $\uparrow$ events ($\uparrow = L\uparrow + U\uparrow$). For archetype-A, L$\uparrow /  \uparrow$ is nearly 1 i.e. 100\%, with a negligible fraction of U$\uparrow$ for all $\tau_p$. In contrast, in archetype-B, the ratio L$\uparrow /  \uparrow$ decreases with longer $\tau_p$ e.g. at $\tau$$_p$ = 30s $\sim$ 10\% $\uparrow$ events are L$\uparrow$, with $\sim$ 90\% being U$\uparrow$ events. This is because at higher voltages and longer $\tau_p$, the increased switching activity results in additional U$\uparrow$ events. The number of linked-events per pulse on the other hand remains similar for both archetypes (around 10--20\%), as is depicted in Fig. \ref{fgr:Fig_counts}(b). Therefore at all voltages, a similar fraction of the LE induce L$\uparrow$ events.

\begin{figure}[h]
\centering
  \includegraphics[width=9cm]{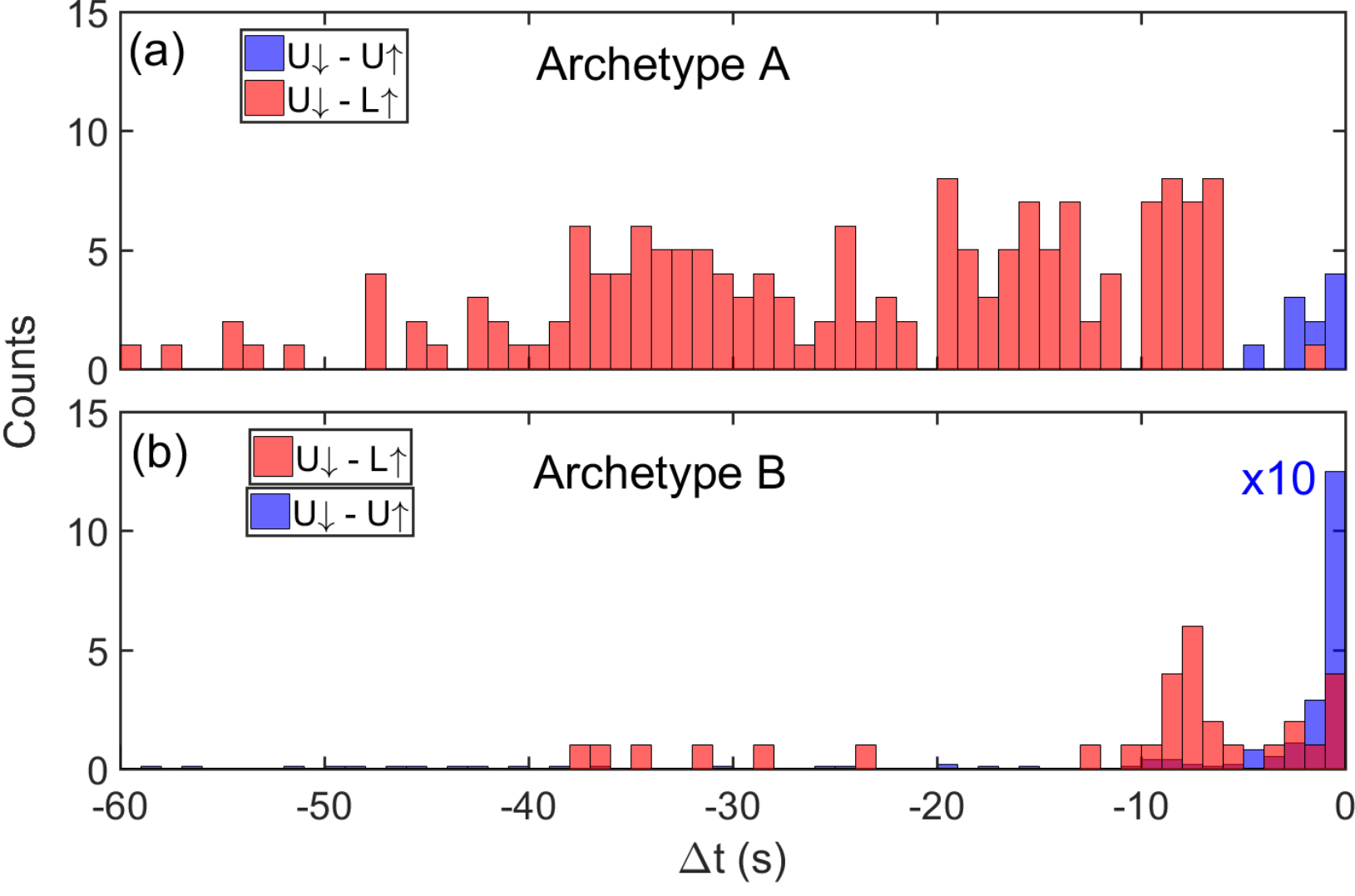}
  \caption{ The event-structures associated with linked-up (L$\uparrow$) and unlinked-up (U$\uparrow$) events for archetype-A and archetype-B. (a) archetype-A shows predominantly (75\%) of the L$\uparrow$ events are preceded by U$\downarrow$ events with few U$\downarrow$-U$\uparrow$ pairs. (b) The situation for archetype-B is different with dominance of U$\uparrow$ (note 10$\times$ scale change), typically preceded by an U$\downarrow$. A much smaller number of L$\uparrow$ events are observed, also typically preceded by U$\downarrow$ .}
  \label{fgr:Fig_HistComp}
\end{figure}

The event-structure associated with both L$\uparrow$ and U$\uparrow$ events across both archetypes is shown in Fig. \ref{fgr:Fig_HistComp}. As can be seen in Fig. \ref{fgr:Fig_HistComp}(a), for archetype-A the L$\uparrow$ events are predominantly (75\%) preceded by a U$\downarrow$ event. The wide distribution of times $\Delta$t between U$\downarrow$ and L$\uparrow$ events is understood to originate from the distribution of electric-field strengths in the gaps in the percolating network\cite{Fostner2014}. The electric field induced formation of atomic wire closes the tunnel gap as shown in Fig. 7 of Ref.\cite{Bose2017}, on time-scales dependent upon both the electric field strength and the position of the synapse in the network. We believe this is similar to an integrate and fire process.\cite{Burkitt2006}

In contrast, for archetype-B the U$\uparrow$ events are more common than L$\uparrow$, and U$\uparrow$ are also predominantly preceded by a U$\downarrow$ with a short $\Delta$t. The narrower distribution of times between the events is due to the higher applied voltage in conjunction with the larger number of active synaptic sites in network. This means that when any one synapse changes state it is likely to cause a change in the state of other synapses, leading to multitude of synapses reconfiguring. Reconfiguration of these interdependent spatially separated synapses via the complex network corresponds to the existence of spatial correlations.

\subsection{Synaptic pathways accessible in archetype A\&B}
\label{synaptic pathways}
As stated in the introduction, all the synaptic elements in the network operate via application of local electric-field (and currents) and are fundamentally similar in character. The relative contribution from individual synapses to the conductance of the network, on the other hand, depends upon their spatial position in the percolating network and thus also determines the  network dynamics.

In archetype-A, i.e. at low voltages, the accessible network is limited to a small number of ``primary" conduction pathways connecting the two electrodes. The applied voltage drops across gaps in these pathways and generates high local electric fields. Thus a small number of available synapses are able to rapidly form atomic-wires resulting in L$\uparrow$ events at the LE (Fig. \ref{fgr:Fig_counts}), with large $\Delta$G [see Fig. \ref{fgr:Fig_Hist3} and associated discussion]. In archetype-B, the synapses on the primary-pathways can still be reconfigured, but the higher voltages also activate additional network-branches. These ``higher order" branches make a smaller contribution towards the overall network conductance and synaptic events on them therefore cause a smaller measured $\Delta$G.

\begin{figure}[h]
\centering
  \includegraphics[width=9cm]{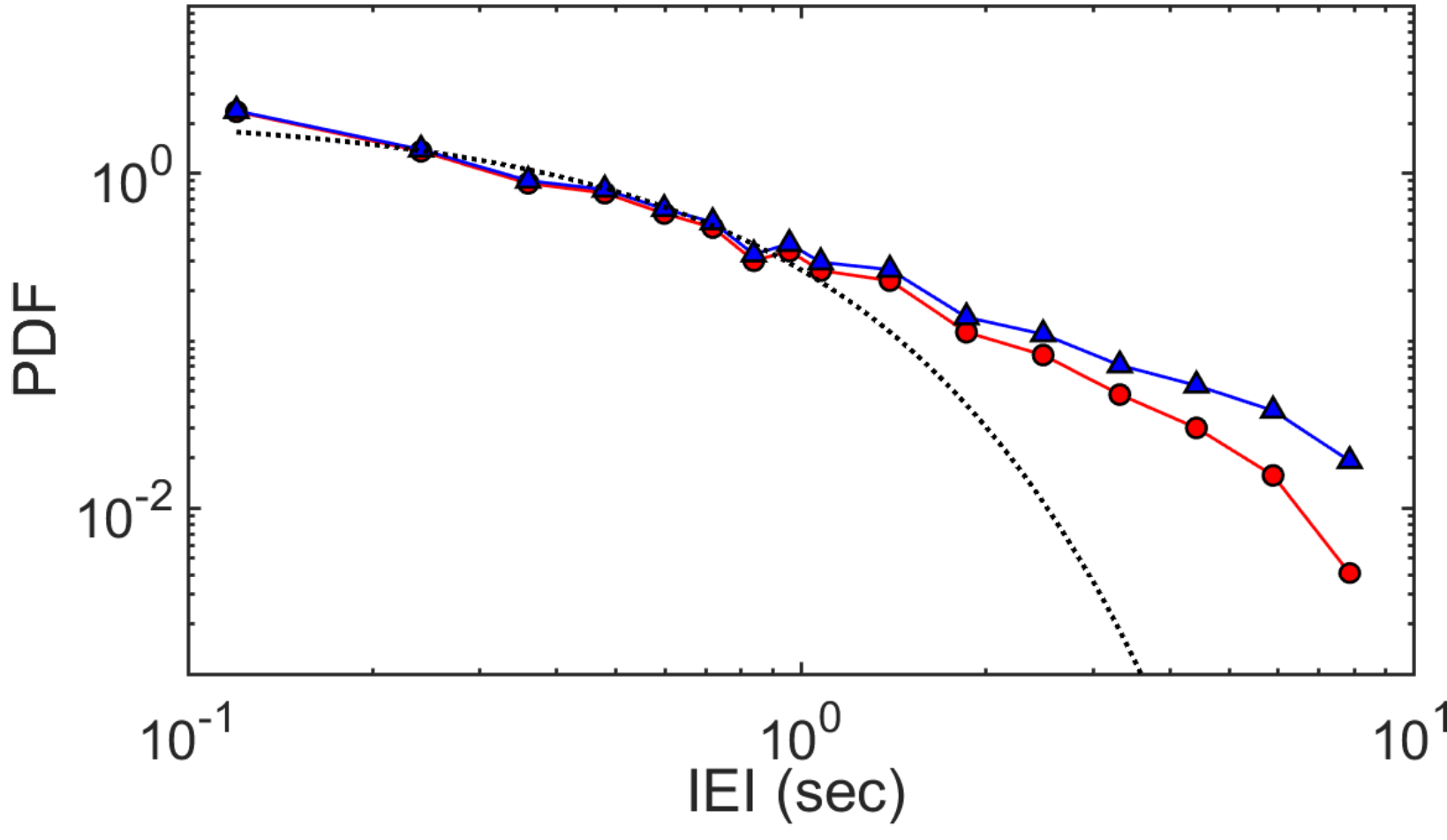}
  \caption{Distribution of inter-event intervals (IEIs) from the device discussed in Figs. \ref{fgr:Fig_ArchABdep}-\ref{fgr:Fig_HistComp}. The raw IEI probability distribution (red circle) shows strong downward curvature at IEI values close to the pulse duration ($\tau$$_p$ = 10s, V$_p =$ 8V) due to the effect of the limited pulse length on long-time correlations. In principle, these effects can be corrected for using the correction factor $1 / (\tau_p - IEI)$  which was applied for $\tau_p$ = 10 s (green triangles). Once this effect is removed, the IEI distribution appears to show a heavier-tail, suggesting temporal correlations.}
  \label{fgr:Fig_powerlaw}
\end{figure}

\subsection{Temporal correlations}
\label{IEI}
To be an effective dynamical reservoir, for example for RC applications, a network must exhibit temporal correlations in addition to the spatial correlations discussed in the previous section.\cite{Stieg2012,Mantas2009,Sillin2013} Indeed it is believed that spatio-temporal correlations are a feature of the biological brain and are a requirement of neuromorphic network architectures in general.\cite{Beggs2003,Stieg2012} Distributions of random, uncorrelated events have statistical distributions of inter-event intervals (IEIs) which decay exponentially. Conversely, correlated non-Poisson processes, where the temporal occurrence of an event depends on the occurrence of previous events, can result in heavy-tailed IEI distributions.\cite{Barabasi1999,Karsai2012}

A probability density function (PDF) of IEIs for a typical device is shown in Fig. \ref{fgr:Fig_powerlaw}. As previously mentioned, there are no synaptic reconfigurations during the off-pulse and so events with IEIs $\geq \tau_p$ are impossible. The raw IEI distribution (red) in Fig. \ref{fgr:Fig_powerlaw} therefore shows strong downward curvature, and at first sight might be mistaken for an exponential decay. However, the dotted line shows clearly that the best fit exponential does not have a sufficiently long tail to account for the experimental distribution: the tail of the experimental distribution is truncated by the finite pulse length. A straightforward calculation leads to a correction factor of $1 / (\tau_p - IEI)$ that can be applied to remove this effect (blue, $\tau$$_p$ = 10s). Even after this correction, the distribution is still slightly curved, and it is not possible to identify the precise functional form of the data.\cite{clauset2009} Nevertheless the distribution is clearly more heavy-tailed than the exponential distribution, providing preliminary evidence of the temporally correlated network activity. Future experiments will explore the spatio-temporal dynamics in the network.

\section{Results II: Network Morphology}
We now discuss the interdependence of the synaptic activity of individual switches and the network morphology.
\subsection{Interrelationship between individual synapses and the network}
\label{morphology}
\begin{figure*}[h]
\centering
  \includegraphics[width=15cm]{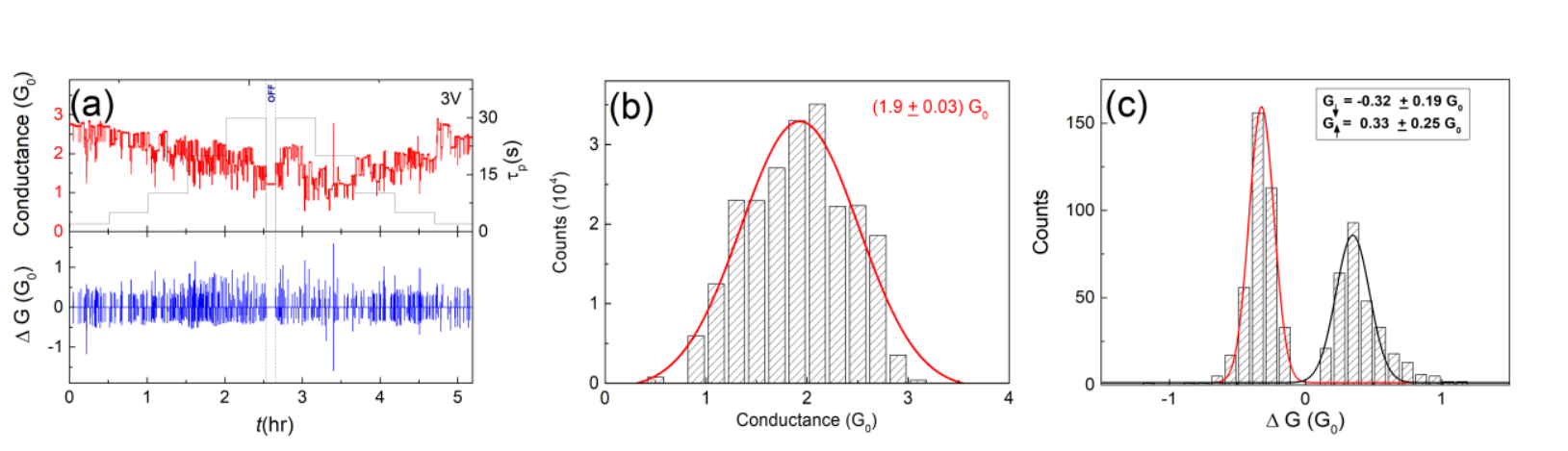}
  \caption{(a) Switching in response to a sequence
of square voltage pulses with fixed $V_p$ = 3V (just above $V_T$)
and variable $\tau_p$ (2 -- 30s). The conductance remains
unaltered for read voltages of 0.1 V as seen in the flat section
(OFF) in middle of the sequence. Bottom panel shows $\Delta G$
for  these switching events. (b) Histogram of the measured initial conductance for
each switching event shows a broad peak around $\sim$2G$_0$. (c)
Histogram of sizes of switching events indicates a clear
preference for $\Delta G = \pm$ 0.3$G_0$. The solid lines in (b)
and (c) are Gaussian fits to the
 distributions. }
  \label{fgr:Fig_Hist1}
\end{figure*}

Detailed analysis of archetype-A switching behavior (V$_p$ = 3V) for a different device is shown in Fig. \ref{fgr:Fig_Hist1}. This data is similar to that shown in
Fig. \ref{fgr:fig_Width1}, and indeed all devices\cite{Bose2017} show similar behaviour.  In
Fig. \ref{fgr:Fig_Hist1}(a), as in Fig. \ref{fgr:fig_Width1}(c, d), longer
$\tau_p$ leads to lower conductance since the electromigration
induced disconnection of atomic wires dominates over the
field-induced reconnection.  Figs. \ref{fgr:Fig_Hist1}(b) and (c) show that histograms of $G$ and $\Delta G$ for each event are
strongly peaked. Importantly, the relatively narrow distributions
indicate that the \emph{individual} switching elements accessible in archetype-A exhibit
well-defined changes in conductance, consistent with the formation
of atomic scale wires\cite{Sattar2013} [if the switching elements
have a large range of conductances, the resulting changes in
conductance of the network must be broadly distributed]. The
non-quantised measured values of $\Delta G$ can be explained by
the series resistance of the percolating film that
surrounds each atomic scale wire -- see further discussion in Section \ref{Gmorph} below.

\begin{figure}[h]
\centering
  \includegraphics[width=6cm]{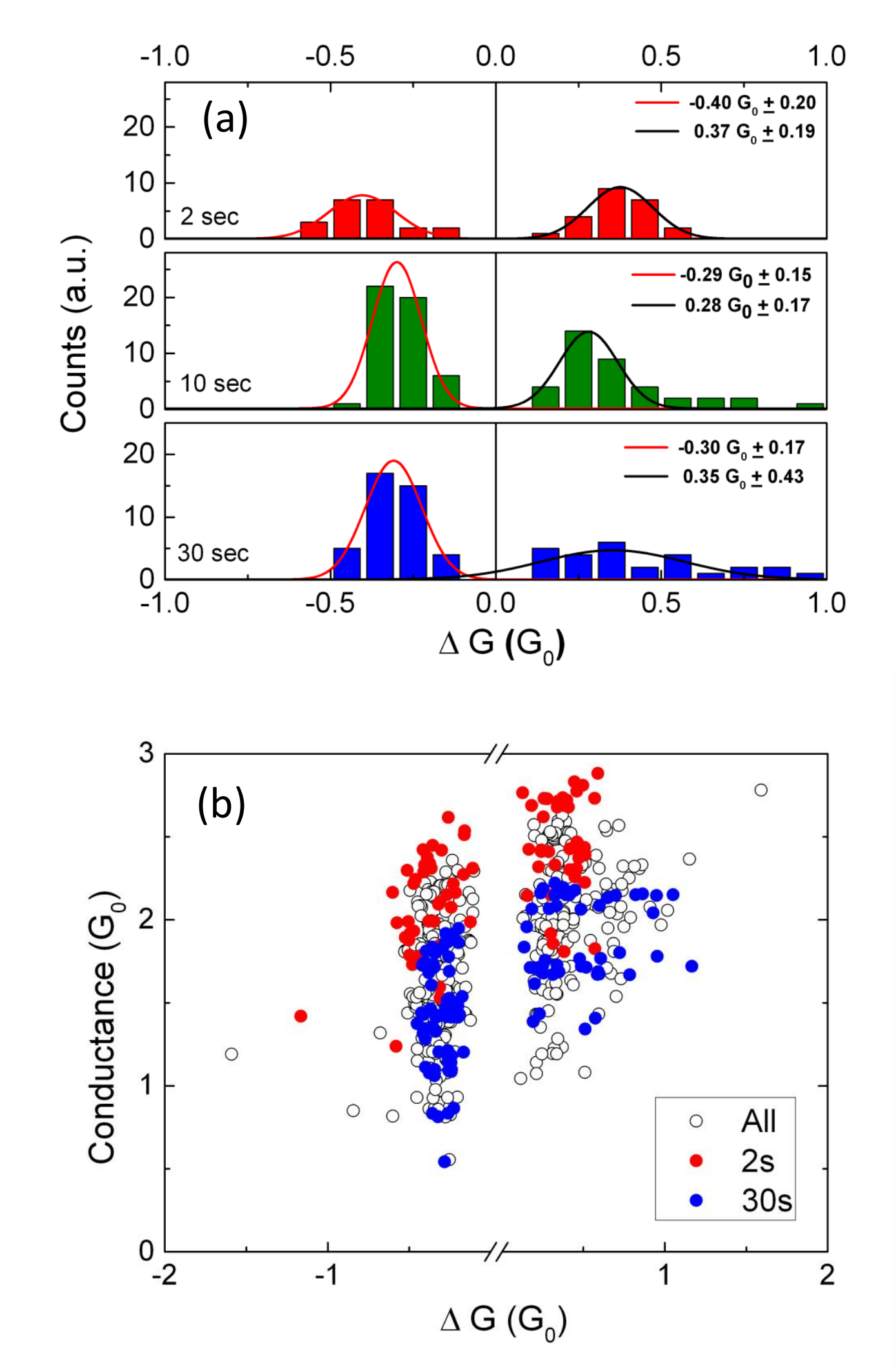}
  \caption{(a) Histograms of switching events for
different $\tau_p$. For shorter pulse widths the majority of
events are centered around $\pm$ 0.3$G_0$. At higher pulse widths
($\tau_p \sim$ 20--30s) the efficient breaking of connections
allows an increase in the number of large $G_{\uparrow}$ events.
(b) Interdependence of $\Delta G$ and $G$ with open circles
representing all events and red and blue filled symbols
representing the subset of the data with $\tau_p$ = 2s and 30s
respectively. The x-axis break indicates that events with $\Delta G
> 0.1 G_0$ are included in the analysis.}
  \label{fgr:Fig_Hist3}
\end{figure}

Fig. \ref{fgr:Fig_Hist3}(a) shows separate histograms of the $\Delta
G$ data for representative $\tau_p$.  $G_{\downarrow}$ events
exhibit peaks for $\Delta G \sim -$0.3$G_0$ whereas $G_{\uparrow}$
events are peaked at $\Delta G\sim$0.3$G_0$ only for small pulse
widths. For longer pulse widths there are a significant number of
large $G_{\uparrow}$ events  and the histogram is not strongly
peaked. This behaviour can be understood as being the result of
the breaking of multiple atomic scale wires during long pulses due
to electromigration. This means that when the next pulse is
applied there are more synapses available for formation of new wires
and larger $\Delta G$  are possible. Fig. \ref{fgr:Fig_Hist3}(b) shows
that $\Delta G$ is not correlated with $G$ for these switching
events, consistent with switching events of the same size (i.e.
quantised events) taking place at multiple sites in the complex
network.

\subsection{Measured conductance changes and device morphology}
\label{Gmorph}
In the present devices long-term stability of the switching
behaviour was achieved by oxidation during
deposition\cite{Bose2017}, and so the resistance of the
paths through the network is non-negligible. In contrast, in Ref. \cite{Sattar2013}
the emphasis was on achieving devices that exhibit large switching
events (large $\Delta G$) and the observation of quantised
resistance ($G = N G_0$), which required a low series resistance.
If an atomic scale wire is formed somewhere within a percolating resistive
network, the rest of the network can be approximated as an
equivalent circuit comprised of a series conductance $\alpha G_0$
and a parallel conductance $\beta G_0$ [$\alpha$ and $\beta$ could
be any real numbers, whereas $N$ is an integer;  all
conductances are in units of $G_0$]. It then requires only a little
algebra  to show that the total conductance of the network is $G =
\alpha \beta / (\alpha + \beta )$ and the change in conductance is $\Delta G  =
\alpha^2  N /[ (\alpha + \beta )(\alpha + \beta + N)]$.

Fig. \ref{fgr:Fig_Hist1} shows that, for this particular device and archetype-A behavior near $V_{th}$, $\Delta G \sim$ 1/3
\emph{and} $G \sim$ 2. Although the positions of the peaks in the histograms of $\Delta G$ and $G$ vary to some extent from device to device, such well defined peaks are commonly observed for archetype-A. These peaks are visible because the voltage (corresponding to an electric field in the gaps in the network) is only sufficient to activate a small number of switching sites (synapses). Higher voltages activate more switching sites (Figs. \ref{fgr:fig_Width1}(e)) and histograms of $\Delta G$ exhibit power law behaviour as complex inter-dependent pathways through the network are activated -- this will be discussed in detail elsewhere.

The results in Fig. \ref{fgr:Fig_Hist1} put strong constraints on
the values of ($\alpha$, $\beta$ , $N$) that can provide solutions
of the above equations. We find numerically that when $N$ is
restricted to physically reasonable\cite{Olesen1994} values less than 10, the values of $\alpha$ and $\beta$ that meet
these constraints are limited to the range $2 < \alpha < 5.5$ and
$\beta < 10$. These solutions indicate that in order to achieve
the experimental results the parallel conductance cannot be too
high, and the series conductance is restricted to a relatively
narrow range. The range of $\alpha$ corresponds to series
resistances $\sim$ 2 - 6 k$\Omega$, while the parallel resistance
must be bigger than $\sim$ 1 k$\Omega$. The measured resistances
(1-10k$\Omega$) of the percolating networks that exhibit switching
effects are consistent with these ranges.

$\alpha$ and $\beta$ represent the electrical properties of the
percolating network and so it follows that the morphology of the
actual device must be such that it provides the required series
and parallel resistances. The sizes of the nanoparticles and
morphology of the network are controlled by coalescence during the
deposition process\cite{Bose2017}, and in light of this analysis
it is clear that one of the functions of the deliberate oxidation
process during deposition of the nanoparticles  is to constrain
the morphology of the network so as to allow observation of
consistent switching behaviour.

\section{Discussion and Conclusions}
Our films of nanoparticles exhibit
complex patterns of electrical switching behaviour at sites corresponding to tunnel gaps within the  percolating network (`synapses'). Voltage  pulses were used to control
formation and annihilation of atomic scale wires at these sites,
thereby modifying the device conductance. Shorter (longer) voltage pulses
bias the network towards higher (lower) conductance as more (fewer) connections are formed across the network.

\begin{figure}[h]
\centering
  \includegraphics[height=4cm]{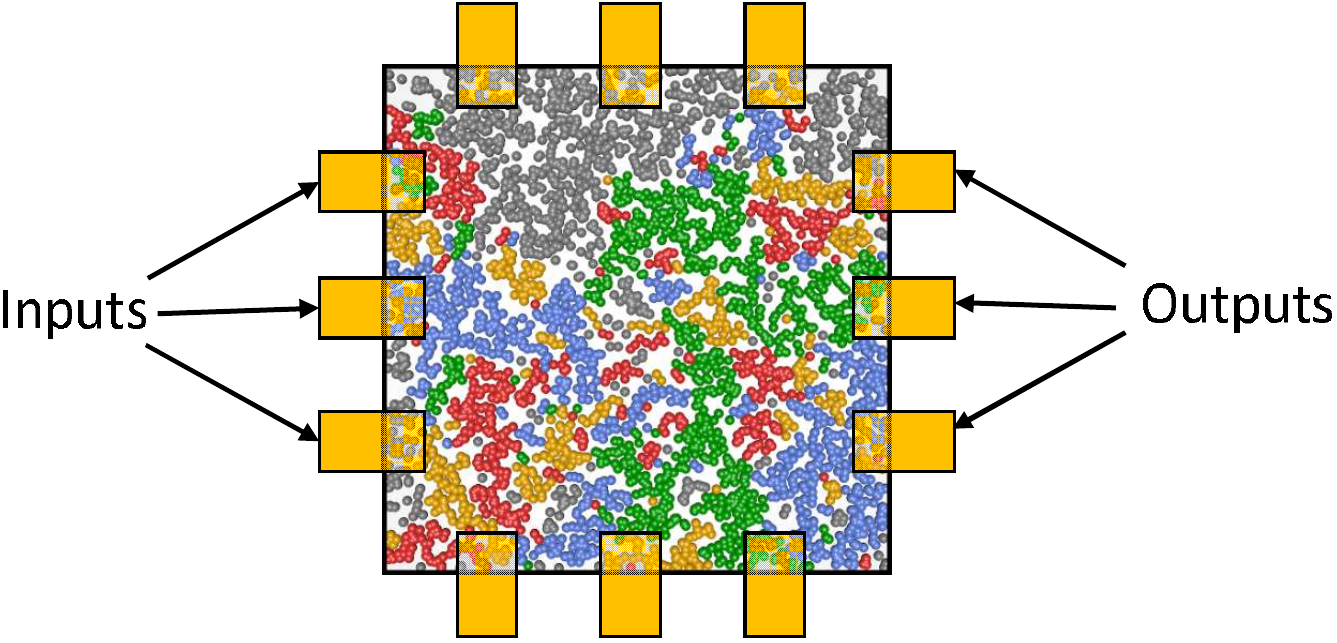}
\caption{\label{fig_chip_2} Schematic of the percolating network
comprising of many groups of particles separated by tunnelling
gaps surrounded by an array of contacts/electrodes which can be
designated as input or output terminals depending on the
computation task at hand.}
\end{figure}

Even in the low $V_p$ regime, switching at one site is influenced by the states of the other synapses. This provides a first indication of the possibility that these percolating networks might exhibit the spatio-temporal correlations required for neuromorphic computing. As the applied voltage is increased, the switching dynamics become more complex because more synapses in the network are activated. We have explored the inter-relationships between the switching events that cause increases and decreases in conductance, and have shown that measured distributions of inter-event intervals are suggestive of the heavy-tailed distributions that are believed to reflect correlated behaviour.\cite{Barabasi1999,Karsai2012}

\subsection{Applications}
The atomic scale processes appear to be similar to
those that occur in Ag/Ag$_2$S atomic switches, which allow
synaptic learning\cite{Ohno2011} and can potentially function on
timescales of $<$ 10ns\cite{Terabe2005}.
The exploitation of this type of synaptic behaviour in self-organised
networks  is still in its infancy, but complex dynamics and spatio-temporal correlations  are essential for development of neuromorphic applications. Such
applications include, for example, spike
timing dependent plasticity \cite{Ohno2011}, learning
behaviour\cite{Fostner2015,Serb2016, Pantazi2016}, rate and population coding \cite{Mizrahi2018}, detection of
temporal correlations\cite{Tuma2016,Querlioz2013} and pattern
recognition\cite{Kulkarni2012, Stieg2012,Service2014, Torrejon2017}.
It is clearly important
now to explore the ultimate limits of the present switching
processes and develop multi-contact devices that would allow demonstration of applications such as pattern recognition.

\subsection{Prospects for further device development}
One of the attractive features of the present networks is the simplicity of the device fabrication process\cite{Bose2017} and the possibility of straightforward extension to a range of more complex devices. Fig.~\ref{fig_chip_2} shows a schematic of the percolating
network, comprising many groups of particles separated by
tunnelling gaps, surrounded by an array of contacts/electrodes
which can be designated as input or output terminals. In order to be able to implement (for example) RC we envisage that initially the  output signals will be recorded and linear regression tasks performed in software, but that in a future generation of devices the electrodes
will provide connections between the percolating network and CMOS circuitry on the same chip, allowing implementation of the required algorithms in hardware.
Various other configurations are of course possible where contacts can be used as gate electrodes and used to modulate the percolating network.
This could be equivalent to modifying the synaptic weights in the interior of an
artificial neural network.

\section*{Conflicts of interest}
There are no conflicts to declare.

\section*{Acknowledgements}
The authors gratefully acknowledge financial support from the Ministry of Business Innovation and Employment, New Zealand, the
Marsden Fund, New Zealand, and the MacDiarmid Institute for
Advanced Materials and Nanotechnology.

\bibliographystyle{rsc} 

\providecommand*{\mcitethebibliography}{\thebibliography}
\csname @ifundefined\endcsname{endmcitethebibliography}
{\let\endmcitethebibliography\endthebibliography}{}

\end{document}